\documentstyle[11pt,epsfig]{article}
\begin{document}
\newcommand{\roughly}[1]%


\newcommand{\PSbox}[3]{\mbox{\rule{0in}{#3}

\includegraphics{#1}\hspace{#2}}}
\newcommand\lsim{\roughly{<}}
\newcommand\gsim{\roughly{>}}
\newcommand\CL{{\cal L}}
\newcommand\CO{{\cal O}}
\newcommand\half{\frac{1}{2}}
\newcommand\beq{\begin{eqnarray}}
\newcommand\eeq{\end{eqnarray}}
\newcommand\eqn[1]{\label{eq:#1}}
\newcommand\intg{\int\,\sqrt{-g}\,}
\newcommand\eq[1]{eq. (\ref{eq:#1})}
\newcommand\meN[1]{\langle N \vert #1 \vert N \rangle}
\newcommand\meNi[1]{\langle N_i \vert #1 \vert N_i \rangle}
\newcommand\mep[1]{\langle p \vert #1 \vert p \rangle}
\newcommand\men[1]{\langle n \vert #1 \vert n \rangle}
\newcommand\mea[1]{\langle A \vert #1 \vert A \rangle}
\newcommand\bi{\begin{itemize}}
\newcommand\ei{\end{itemize}}
\newcommand\be{\begin{equation}}
\newcommand\ee{\end{equation}}
\newcommand\bea{\begin{eqnarray}}
\newcommand\eea{\end{eqnarray}}

\def\Dsl{\,\raise.15ex \hbox{/}\mkern-12.8mu D}
\newcommand\Tr{{\rm Tr\,}}
\thispagestyle{empty}

\begin{titlepage}
\begin{flushright}

CALT-68-2415\\

\end{flushright}

\vspace{1.0cm}

\begin{center}

{\LARGE \bf  Debt Subordination and The Pricing of Credit Default Swaps }\\

\bigskip

\bigskip

\bigskip

{ Peter B. Lee$^a$,  Mark B. Wise$^b$ and Vineer Bhansali$^c$} \\

~\\

\noindent

{\it\ignorespaces

          (a) California Institute of Technology, Pasadena CA 91125\\

           {\tt peter@theory.caltech.edu}\\

\bigskip (b) California Institute of Technology, Pasadena CA 91125\\

          {\tt wise@theory.caltech.edu}\\

\bigskip   (c) PIMCO, 840 Newport Center Drive, Suite 300\\

               Newport Beach, CA 92660 \\

{\tt   bhansali@pimco.com}

}\bigskip

\end{center}

\vspace{1cm}

\begin{abstract}
First passage models, where corporate assets undergo a random walk
and default occurs if the assets fall below a threshold, provide
an attractive framework for modeling the default process. Recently
such models have been generalized to allow a fluctuating default
threshold or equivalently a fluctuating total recovery fraction
$R$. For a given company a particular type of debt has a recovery
fraction $R_i$ that is greater or less than $R$ depending on its
level of subordination. In general the $R_i$ are functions of $R$
and since, in models with a fluctuating default threshold, the
probability of default depends on $R$ there are correlations
between the recovery fractions $R_i$ and the probability of
default. We find, using a simple scenario where debt of type $i$
is subordinate to debt of type $i-1$, the functional dependence
$R_i(R)$ and explore how correlations between the default
probability and the recovery fractions $R_i(R)$ influence the par
spreads for credit default swaps.  This scenario captures the
effect of debt cushion on recovery fractions.
\end{abstract}

\vfill


\end{titlepage}


\section{Introduction}

The Merton framework [Merton (1974)] provides an attractive
approach to credit risk, relating default probabilities for
corporations to their stock prices. The default process is usually
modeled by assuming that the corporate assets undergo a simple
random walk and the first time the assets fall below a threshold
$T$ default occurs [Black and Cox (1976), Longstaff and Schwartz
(1995), Leland and Toft (1996), {\it etc}.]. A company may have
several types of debt, type $i$ contributing $D_i$ to a total debt
$D$ ($\sum_i D_i=D$). The debt holders weighted, by debt fraction,
average recovery fraction (i.e., total recovery fraction) $R$ is equal to $T/D$. If default
occurred when the assets first passed a threshold equal to $D$ then
the total recovery fraction would be one. However, the total recovery
fraction is typically considerably less than one and its precise
value is not known until after the default actually occurs. Hence
the default threshold $T$ or equivalently the total recovery fraction
$R$ should be treated as a random variable. Recently first passage
models have been generalized to include a fluctuating recovery
[Pan (2001), Finkelstein, et. al. (2002)].  Expected default probabilities
and par spreads $c_i$ for credit default swaps have been computed
using a first passage model with a fluctuating total recovery
fraction $R$. The expected default probability, $P_D(t)$, for example,
is calculated by taking the usual first passage time expression $P_D(t|R)$
and taking its expected value over possible values for $R$.

Since the recovery fraction for the $i$'th type of debt, $R_i$ is
not equal to $R$ it is usually treated as a fixed independent
quantity in the calculation of par spread, $c_i$ for a default
swap on corporate bonds which form the $i$'th debt
type\footnote{For an excellent discussion on the treatment of recovery
in various classes of models for credit risk see Altman et. al. (2002).}. This
assumption neglects the empirical observation that recovery fractions
for different classes of bonds typically depend strongly on the
total recovery, and there is considerable uncertainty regarding
the expected recovery for each class. We argue that the $R_i$'s
are actually functions of $R$ and the main purpose of this paper
is to study the implications of this relationship for the pricing
of credit default swaps on different subordination classes of
bonds. Clearly the $R_i$ cannot be completely independent of $R$.
For example, since
\be
 \sum_i R_i D_i =R D,
\ee
if $R$ and the $R_i$ go between zero and one then $R=0$ implies that
all of the $R_i$ are zero and $R=1$ then all of the $R_i=1$.

In a simple scenario where debt of type $i$ is subordinate to debt of
type $i-1$ we find the functional dependence of the $R_i$ on $R$
and use it to compute the par credit default swap spreads, $c_i$.
Since the default probability $P_D(t|R)$ depends on $R$ and the recovery
fractions $R_i$ now also depend on $R$ there are correlations
between these two quantities which play an important role. We
examine the significance of these correlations and find that they
cause a substantial decrease in default swap par spreads.  The
effect is very dramatic for the most senior debt.

Debt cushion for debt of type $i$ refers to the proportion of
total debt occupied by those junior to it [Van de Castle and
Keisman (1999)]. The simple scenario we discuss in this paper
relates recovery fractions $R_i$ to $R$ in a way that incorporates
the impact of debt cushion on the pricing of credit default swap
spreads and we find that expected recovery rates increase substantially as the
debt cushion increases.

\section{Credit Default Swaps in a First Passage Model With Fluctuating Default Threshold}

We take the corporate assets $V_t$ to undergo the Ito process,
\be
 {dV_t \over V_t}=\sigma dW_t +\mu dt,
\ee
where,
\be
 E\left[(dW_t)^2\right]=dt
\ee
and $\mu$ is the coefficient of a drift term.  The stochastic
process for $\log(V_t)$ is obtained using Ito's lemma
 \be
   d\left(\log V_t\right) = \sigma dW_t + \left(\mu-{\sigma^2\over
   2}\right) dt,
 \ee
where $\log$ denotes the natural logarithm.  Integrating over time
 \be
  V_t=V_0\exp\left(\sigma W_t+ \left(\mu-{\sigma^2\over 2}\right) t\right).
 \ee
In this paper, we set $\mu=0$ so that the expected value, $E[V_t]$
is independent of time [Pan (2001)].\footnote{For the purpose of pricing
credit default swaps, one should use the risk-neutral measure $dV = r V_t dt - C dt +
\sigma dW_t$ if the answer to the delicate question of $V_t$ being a traded asset
is affirmative.  Here, $r$ is the risk-free interest rate and $C$ arises from dividends
and interest payments made on the debt.  In this article, we drop the drift terms $r$ and $C$
in order to concentrate on the effects of debt subordination and debt cushion on the pricing of CDS.}
Default occurs if the assets $V_t$ fall below a threshold $T$. If the total corporate debt is $D$,
the average recovery fraction weighted by debt fraction (i.e., total recovery fraction) is $R=T/D$.
Because the total recovery fraction is not known at the initial time, the default threshold $T$ or
equivalently the total recovery fraction $R$ is taken to be a random variable
with probability distribution $P(R)$. The expected survival probability,
$P_S(t)$, for the fluctuating default threshold case results from
weighting the standard (i.e., fixed default threshold) first
passage time expression with the probability distribution $P(R)$.
The survival condition is $V_t > T$ which implies that,
 \beq
  W_t-{\sigma \over 2} t > {1\over \sigma} \log\left({RD\over V_0}\right).
 \eeq
The survival probability can be deduced from the above condition
since it simply leads to a constant drift Brownian motion with an
absorbing barrier. Introducing the convenient notation to signify
expectations of a quantity $f(R)$ over the recovery probability
density function
 \be
  \label{average}
  \langle f(R)\rangle=\int_0^1 dR P(R)f(R),
 \ee
the expected survival probability is
 \be
  P_S(t)=\langle P_S(t|R)\rangle
 \ee
where
 \be
  \label{key}
  P_S(t|R)=\Phi \left(-{B(R) \over {\sqrt t}}- {\sigma {\sqrt t} \over
  2} \right)-\exp\left( {- \sigma B(R)}\right)\Phi
  \left({B(R) \over {\sqrt t}} - {\sigma {\sqrt t} \over 2}
  \right).
 \ee
In equation (\ref{key})
 \be
  B(R)=-{1 \over \sigma}\log \left( {V_0 \over DR} \right)
 \ee
and $\Phi$ is the cumulative normal distribution,
 \be
  \Phi(a)={1 \over \sqrt{2 \pi}} \int_{-\infty}^a dx \exp\left( -{x^2 \over 2} \right).
 \ee

Although there have been reports of recovery
rates greater than unity, such events are rare and untypical.
Hence, we take the upper limit in the range of integration in
equation (\ref{average}) to be $R=1$. When $V_0/D$ is significantly greater than unity
this causes an unrealistically large suppression of default probabilities at small times.
It is straightforward to extend the results presented here to allow for $R>1$.

The credit default swap par spread $c_i$ for a particular type of
debt labelled by $i$ can be found by equating the expected loss due
to the firm defaulting with the credit default swap
payments, which are assumed to be made
continuously\footnote{Equation (\ref{coupon}) can be easily
generalized to spread payments that are made discretely.}. We have
 \beq
  \label{coupon}
  c_i = {\left\langle \left(1-R_i(R)\right) \left(1-P_S(0|R)- \int_0^t ds
  {\partial P_S(s|R)\over \partial s} e^{-\hat r(s) s} \right)  \right\rangle \over
  \left\langle \int_0^t ds P_S(s|R) e^{-\hat r(s) s} \right\rangle},
 \eeq
where $\hat r(s)$ is the spot rate of term $s$.

In the remainder of this section, we consider a simple
scenario for the recovery rates pertinent to different classes of
debt at the point of default. Consider $N$ different classes of
debts
 \beq
  D_i, \;\;\;\;\; 1 \le i \le N
 \eeq
in the descending order of seniority such that $D_1$ is the most
senior.  We have the following constraints on the debts:
 \bea
  && \sum_{i=1}^N D_i = D, \\
  && \sum_{i=1}^N R_i D_i = R D.
 \eea
Under the current U.S. bankruptcy legislation, assets of a
bankrupt firm are distributed to its creditors according to the
{\it Absolute Priority Rule} [Gupton and Stein (2002)].  Senior
debt holders are able to have a claim on the remaining assets
before the junior holders can do so. However, in practice the
actual recovery rates may depend on a ``plan" agreed upon by the
claimants.  In this article, we apply the Absolute Priority Rule
in its strictest sense.  We assume that senior holders, when given
the chance, are able to claim up to the full recovery rate at
which point the next senior holder is able to stake a claim. For
example, the most senior holder's recovery rate initially goes as
$R_1=RD/D_1$ then caps at $1$. The next senior debt holder is
unable to recover any amount of the asset until $R=D_1/D$, then he
recovers $(R D-D_1)/D_2$ until the recovery rate caps at $1$.
Repeating this argument, one can show that for the $i$'th
subordinate, the recovery fraction can be expressed as
 \bea
  && R_i(R) =  \alpha_i(R) \left[
  \Theta\left(\alpha_i(R)\right) -
  \Theta\left(\alpha_i(R)-1\right)\right]
   + \Theta\left(\alpha_i(R)-1\right), \\ \nonumber
  && \alpha_i(R) = {D\over D_i} \left(R - \sum_{j=1}^{i-1} (D_j/D)\right),
 \eea
where  $1 \le i \le N$ and $R\in[0,1]$.  $\Theta$ is the step
function defined in the following manner:
 \bea
  \Theta(x) = \left\{ \begin{array}{ll}
                        1 & \mbox{if $x\ge 0$} \\
                        0 & \mbox{otherwise.}
                        \end{array}
                        \right.
 \eea
Obviously, the general expression for the credit default swap par
spread $c_i$ given in equation (\ref{coupon}) can accommodate any
form of the recovery rate as a function of the total recovery rate
on total debt. The scenario we consider with the recovery rates
depending only on the seniority of the debt and the level of debt
cushion should prove useful in a general framework for pricing
credit default swaps.

\section{Numerical Results}
\begin{figure}[htb]
\begin{center}
\epsfxsize=3.1in\leavevmode\epsfbox{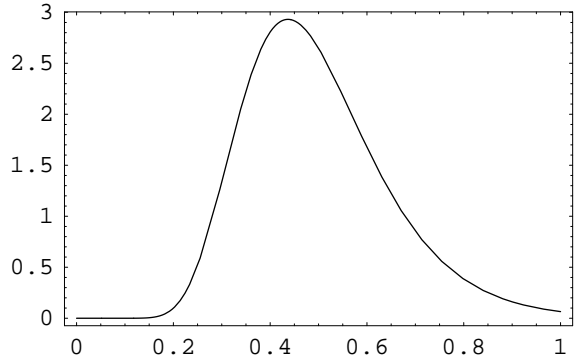}
\end{center}
\caption{$P(R)$ from recovery distribution for years 1987
to 1997.}
\end{figure}

The total recovery rate for a company is correlated with the level of
default activity in the market and is dependent on numerous
macroeconomic as well as company specific factors.  In this paper,
we only concentrate on the effect of seniority on the recovery
rates $R_i$. We take the total recovery rate
to follow the lognormal probability distribution
\beq
\label{probr}
P(R) = {6.48933 \times 10^{-2} \over R^{9.20164}}\exp\left(-{50 \over 9}(\log(R))^2\right),
\eeq
which was fit to recovery data from non-financial firms
that defaulted from 1987 to 1997 [Finkelstein, et. al. (2002)]. Figure 1 plots $P(R)$ in
equation (\ref{probr}) as a function of $R$.

We consider a firm with initial asset to debt ratio of $V_0/D =2$,
asset volatility $\sigma=0.4$ and capital structure composed of
three types of debt with, $D_1/D=0.5$, $D_2/D=0.1$ and
$D_3/D=0.4$.  We have chosen the debt cushion for debt of type $1$
to be $50\%$ so that its average recovery rate of $88\%$ is
significantly less than unity. For example, a debt cushion of $75\%$
implies an average recovery rate $\langle R_1(R)\rangle$ of nearly
$100\%$ and a $25\%$ debt cushion yields average recovery rate of $65\%$.
If we identify debt of type $1$ with bank loans, these average
recovery rates are in reasonable agreement with the results reported by
Van de Castle and Keisman (1999).

\begin{figure}[htb]
\begin{center}
\epsfxsize=3.2in\leavevmode\epsfbox{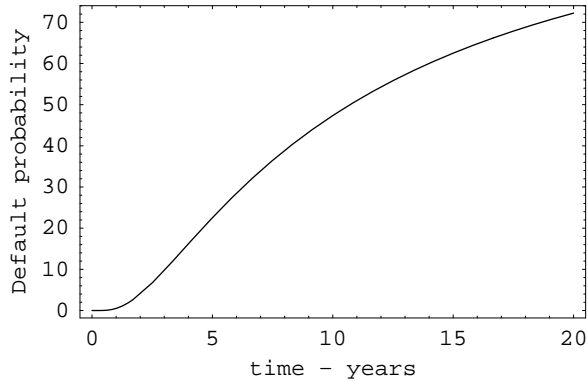}
\end{center}
\caption{Cumulative default probability (in percent) versus time
(in years) with $P(R)$ given in Figure 1  and parameters
$V_0/D=2$, $\sigma=0.4$.}
\end{figure}
\begin{figure}[htb]
\begin{center}
\epsfxsize=3.2in\leavevmode\epsfbox{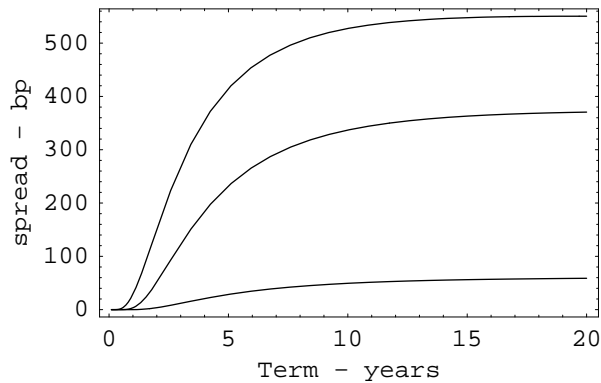}
\end{center}
\caption{Annualized credit default spreads (in bps) versus
maturity (in years) for $D_1/D=0.5$, $D_2/D=0.1$, $D_3/D=0.4$,
$V_0/D=2$, $\hat r(s)=0.05$ and $\sigma=0.4$.}
\end{figure}
\begin{figure}[htb]
\begin{center}
\epsfxsize=3.2in\leavevmode\epsfbox{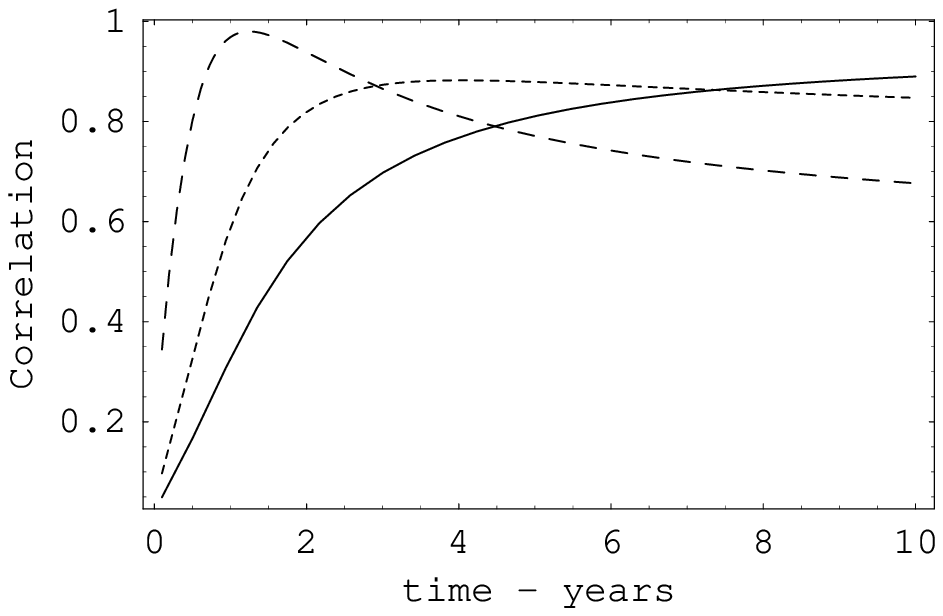}
\end{center}
\caption{Correlations $\xi_i(t)$ of cumulative default probability and the
recovery rates $R_i$.  The solid curve is for $i=1$, short dashed $i=2$ and the
long dashed is for $i=3$.}
\end{figure}

Using $P(R)$ shown in Figure 1 its cumulative expected default
probability, $P_D(t)=1-P_S(t)$, is plotted in Figure 2 as a
function of time $t$. Note the suppression at small times. For
example, $P_D(1{\rm yr})$ is only $0.5\%$. As we remarked earlier
this occurs because $V_0/D$ is significantly greater than unity
and fluctuations of $R$ greater than $1$ are forbidden.

The effect of debt subordination on the pricing of credit
default swaps for this firm's debt is illustrated in
Figure 3, where the par spreads, $c_i$ are plotted as a function of debt maturity.
The lower curve is for debt of type $1$, middle
curve for debt of type $2$ and the upper curve is for the least senior debt of type $3$.
\begin{figure}[htb]
\begin{center}
\epsfxsize=3.2in\leavevmode\epsfbox{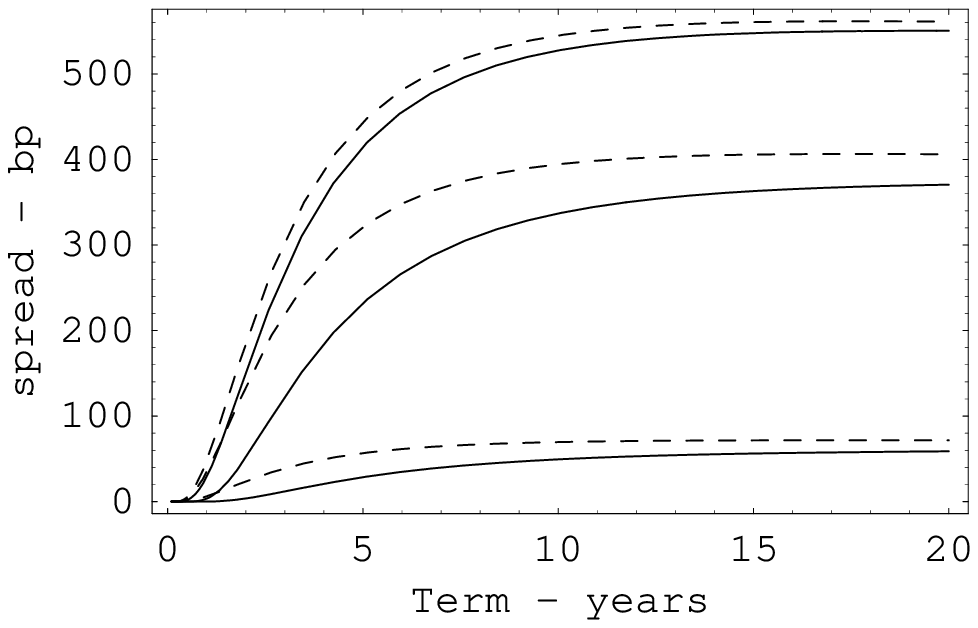}
\end{center}
\caption{Credit default spreads (in bps) versus maturity (in
years) for $\langle R_1\rangle=0.88$, $\langle R_2 \rangle =
0.32$, $\langle R_3 \rangle=0.06$ with the same parameters as in
Figure 1. Solid curve is with correlation and dotted curve is
without.}
\end{figure}

 The correlations between recovery the rates $R_i(R)$ and default probability $P_D(t|R)=1-P_S(t|R)$ are defined by
\beq
\xi_i(t)={\langle R_i(R)P_D(t|R) \rangle-\langle R_i(R)\rangle \langle P_D(t|R) \rangle \over \sqrt{ \langle R_i(R)^2 \rangle - \langle R_i(R) \rangle ^2} \sqrt{ \langle P_D(t|R)^2 \rangle - \langle P_D(t|R) \rangle ^2}},
\eeq
and they are plotted in Figure 4 as a function of time. The correlations are positive
since both the cumulative default probability $P_D(t|R)$ and the recovery fractions
$R_i(R)$ are increasing functions of total recovery fraction $R$. These strong correlations
lead to a substantial reduction in par spreads for credit default swaps.

In Figure 5, we demonstrate the effect of including the correlation
between $R_i(R)$ and $P_D(t|R)$ on the pricing of credit default swaps.  The solid curve plots the
$c_i$'s with the correlation included and the dotted curve plots
the $c_i$'s with $R_i(R)$ in equation (\ref{coupon}) set to the
constant expected recovery value $\langle R_i(R) \rangle$. The
correlation substantially decreases credit default par spread
values. For example, with the parameters we have chosen the
cumulative $ 5 {\rm yr}$ default probability for the company is
$23\%$ and the par spread for a $5$ yr credit default swap on the
type 2 debt is $c_2=232{\rm bp}$. However if $R_2(R)$ is replaced
by $\langle R_2(R)\rangle=32\%$ in equation (\ref{coupon}), then
one finds $c_2=322{\rm bp}$.

The effect of this correlation is more significant for the most
senior type 1 debt. In that case the recovery is one unless $R$
has fluctuated below $0.5$. The credit spread then only gets a
contribution from this region of $R$ integration in equation
(\ref{coupon}). But for $R$ in this region the default probability
is small resulting in a small credit default swap par
spread. For example, for a $5 {\rm yr}$ maturity default swap we
find that $c_1$ is only  $29 \rm bp$, however if $R_1(R)$ is
replaced by $\langle R_1(R)\rangle=88\%$ in equation
(\ref{coupon}) then one finds the much larger value
$c_1=57{\rm bp}$. The very large suppression of credit spreads for the most
senior debt caused by the correlation between $R_i(R)$ and $P_D(t|R)$ may indicate
a problem with the general picture where corporate assets
undergo a simple random walk in time, default occurs when the asset value first crosses
a threshold, the uncertainty of recovery is associated with universal fluctuations in the
default threshold and credit default swap spreads are computed by equating expected cash flows.

\section{Conclusion}

Within the  context of a first passage default model with a
fluctuating default threshold we have explored the implications of
the correlation between the recovery fraction for a particular
type of debt $R_i(R)$ and the default probability $P_D(t|R)$ for the pricing of
credit default swaps.

A fluctuating default threshold is equivalent to a fluctuating
total recovery rate $R$. In general a company has several types of
debt $D_i$ each with a different recovery rate $R_i$ and the total
recovery rate satisfies $R D= \sum_i R_i D_i$, where $D$ is the
total debt. In a simple scenario where debt of type $i$ is
subordinate to debt of type $i-1$, we explicitly derived the
functional relationship $R_i(R)$.  The impact of debt cushion on
recovery rate has been known for some time, and this scenario captures
its effect on the pricing of credit default swap spreads.
Using a form for the probability of total recovery
$P(R)$ suggested by historical data, we calculated par spreads
$c_i$ for credit default swaps on corporate bonds with different
levels of subordination. We found that the correlation between
$R_i(R)$ and $P_D(t|R)$ dramatically decreases the par spreads for
credit default swaps and this effect is greatest for the most
senior debt.

As Figure 4 shows there is a positive correlation between recovery rates
$R_i(R)$ and default probabilities $P_D(t|R)$ since both these quantities are increasing functions
of the total recovery fraction $R$. There is mounting
empirical evidence for a negative correlation between default
probabilities and recovery rates [Frye (2000a), Frye (2000b),
Altman (2001), Carey and Gordy (2001), Hamilton et. al. (2001), and
Altman and Brady (2002)]. However, it is possible to accommodate
this empirically observed behavior within the general class of
models we are discussing. For example, the default probability
increases as $\sigma$ does, and one could introduce a
correlation where companies with a larger asset volatility
$\sigma$ have a lower mean default threshold.  Even in such an extension of the
model explicitly discussed in this paper the functional dependence
of the recovery rates for particular types of debt $R_i$ on the
total recovery rate (on the total debt) $R$ will play an important
role in the correct pricing of credit default swaps.

\vskip1.0in
\noindent{\Large{\bf References}}
\vskip0.25in

\noindent Altman (2001), {\it Altman High Yield Bond and Default Study}, Solomon Smith Barney, U.S. Fixed Income High Yield Report, July. \vspace{0.2cm}

\noindent Altman, E. and Brady, B. (2002), {\it Explaining Aggregate Recovery Rates on Corporate Bond Defaults}, NYU Salomon Center Working Paper Series. \vspace{0.2cm}

\noindent Altman, E., Brady, B., Resti, A. and Sironi, A. (2002) {\it The Link between Default and Recovery Rates: Implications for Credit Risk Models and Procyclicality}, working paper. \vspace{0.2cm}

\noindent Black, F. and Cox, J. (1976) {\it Valuing Corporate
Securities: Some Effects of Bond Indenture Provisions}, Journal of
Finance, 31, 351-367. \vspace{0.2cm}

\noindent Carey, M. and Gordy, M. (2001) {\it Systematic Risk in Recoveries on Defaulted Debt}, working paper presented at the 2001 Financial Management Association Meetings, Toronto, October 20. \vspace{0.2cm}

\noindent Finkelstein, V., Lardy, J.P., Pan, G., Ta, T. and
Tierney, J. (2002), Credit Grades Technical Document, edited by C.
Finger. \vspace{0.2cm}

\noindent Frye, J. (2000a) {\it Collateral Damage Detected}, Federal Reserve Bank of Chicago, Working Paper, Emerging Issues Series, October 1-14. \vspace{0.2cm}

\noindent Frye, J. (2000b) {\it Depressing Recoveries}, Federal Reserve Bank of Chicago, Risk, January. \vspace{0.2cm}

\noindent Gupton, G. and Stein, R. (2002) {\it LossCalc: Moody's
Model for Predicting Loss Given Default (LGD)}, Moody's Investors
Service, February. \vspace{0.2cm}

\noindent Hamilton, D., Gupton, G., and Berthault, A. (2001) {\it
Default and Recovery Rates of Corporate Bond Issuers: 2000},
Moody's Investment Service, February. \vspace{0.2cm}

\noindent Leland, H. and Toft, K. (1996) {\it Optimal Capital
Structure, Endogenous Bankruptcy and the Term Structure of Credit
Spreads}, Journal of Finance, 51, 987-1019. \vspace{0.2cm}

\noindent Merton, R. (1974), {\it On Pricing of Corporate Debt:
The Risk Structure of Interest Rates}, Journal of Finance 29,
449-470. \vspace{0.2cm}


\noindent Longstaff, F. and Schwartz, E. (1995) {\it A Simple
Approach to Valuing Risky Floating Rate Debt}, Journal of Finance,
50, 789-819. \vspace{0.2cm}

\noindent Pan, G. (2001) {\it Equity to Credit Pricing}, Risk,
99-102, November. \vspace{0.2cm}

\noindent Van de Castle, K. and Keisman, D. (1999) {\it Recovering
Your Money: Insights Into Losses From Defaults}, Standard \&
Poor's CreditWeek, June.
\end{document}